\documentclass[fleqn,usenatbib]{mnras}

\usepackage{newtxtext,newtxmath}

\usepackage[T1]{fontenc}

\DeclareRobustCommand{\VAN}[3]{#2}
\let\VANthebibliography\thebibliography
\def\thebibliography{\DeclareRobustCommand{\VAN}[3]{##3}\VANthebibliography}

\usepackage{graphicx}	
\usepackage{amsmath}	

\usepackage{amssymb}	

\title[The magnetic field of NGC 300 ULX1]{Study on the magnetic field strength of NGC 300 ULX1}

\author[Y. Y. Pan et al.]{
Y. Y. Pan,$^{1}$\thanks{E-mail: panyy@xtu.edu.cn}
Z. S. Li,$^{1}$\thanks{E-mail: lizhaosheng@xtu.edu.cn}
C. M. Zhang$^{2,3,4}$
and J. X. Zhong$^{1}$
\\
$^{1}$Key Laboratory of Stars and Interstellar Medium, Xiangtan University, Hunan 411105, China\\
$^{2}$National Astronomical Observatories, Chinese Academy of Sciences, Beijing 100101, China\\
$^{3}$CAS Key Laboratory of FAST, Chinese Academy of Sciences, Beijing 100101, China\\
$^{4}$School of Physical Sciences, University of Chinese Academy of Sciences, Beijing 101400, China
}

\date{Accepted XXX. Received YYY; in original form ZZZ}

\pubyear{2021}

\begin{document}
\label{firstpage}
\pagerange{\pageref{firstpage}--\pageref{lastpage}}
\maketitle

\begin{abstract}
NGC 300 ULX1 is a pulsating ultraluminous X-ray source (PULX) with the longest spin period of $P\simeq31.6\,\rm s$ and a high spin-up rate of $\dot P\simeq-5.56\times10^{-7}\,\rm s\,s^{-1}$ that is ever seen in the confirmed PULXs.
In this paper, the inferred  magnetic field of NGC 300 ULX1 is $\sim3.0\times10^{14}\,\rm G$ using
the recent observed parameters after its first detection of pulsations. According to the evolved simulation of the magnetic field and the spin period, it will become a recycled pulsar or a millisecond pulsar under the conditions of the companion mass and the accretion rate limitation. We suggest that NGC 300 ULX1 is an accreting magnetar accounting for its super Eddington luminosity.
We also propose that there might be other accreting magnetars in the confirmed PULXs. Such PULXs will be helpful for understanding the magnetar evolution and the millisecond pulsar formation whose magnetic field is stronger than $\sim10^{9}\,\rm G$. 
\end{abstract}

\begin{keywords}
stars: neutron--
pulsars: individual: NGC 300 ULX1--
stars: magnetic fields--
X-rays: binaries.

\end{keywords}



\section{Introduction}

Ultraluminous X-ray sources (ULXs) are the compact objects with the  isotropic X-ray luminosity ($L_{\rm X}$) higher than $10^{39}\,\rm erg\ s^{-1}$. They are usually thought as the accreting black holes (BHs) with the stellar or intermediate mass \citep{mil04, liu08, rob16}.
In 2014, M82 X-2 was confirmed to be the first pulsating ULX (PULX) with the detection of a $1.37\,\rm s$ pulsation. It breaks the usual knowledge that ULXs are BHs and indicates there might be more neutron stars (NSs) rather than BHs in ULX populations  \citep{bac14,eks15, sha15,kin16, erk19, kin20}. 

More than 10 ULXs have been confirmed to be pulsars (PSRs), whose spin-up rate ($ |\dot P|\sim 10^{-7}-10^{-10}\,\rm s\,s^{-1}$) and X-ray luminosity ($L_{\rm X}\sim10^{39}-10^{41}\,\rm erg\,s^{-1}$) are different with the known accreting NSs \citep{kin20, son20, erk20}. 
It is suspected that the extraordinary characteristics of PULXs  can be understanding through their magnetic field.
One interpretation is that PULXs are with fields in the magnetar-level $10^{14}\,\rm G$. Such a strong magnetic field can reduce the electron scattering cross-section and promotes super-Eddington luminosities  \citep{dal15, eks15, ton15, pan16, isr17}. 
The other explanation is that the magnetic fields of PULXs are in the $10^{12}$-- $10^{13}\,\rm G$ range, which matches with the field range of normal pulsars \citep{bac14, fur17, car18, kin20}. 
\citet{erk20} found that both magnetar and submagnetar fields were reasonable ($\sim10^{14}\,\rm G$ and $\sim10^{12}\,\rm G$), that was 
with considering a variety of possible spin and luminosity states for each PULX. 
Moreover, an even low magnetic field ($\sim10^{9}\,\rm G$) of M82 X-2 had been suggested by \citet{klu15}, which allows the accretion disc extended to the stellar surface. 
The ultra luminous of PULXs can be attributed to the existence of some degree anisotropy to the radiation field to see the pulsations. A beaming factor can be introduced into the relation between the accretion and isotropic X-ray luminosities as 
$L_{\rm acc}=bL_{\rm X}$, where $b<1$ \citep{fen11}. However, \citet{mus21} found that most confirmed PULXs were with large pulsed fraction and their pulse profiles were nearly sinusoidal, that implied a moderate geometric beaming affection instead of the strong one.  

NGC 300 ULX1 is the slowest pulsar with the highest spin-up rate that has ever been observed in the family of known PULXs.  According to the observations of {\em NuSTAR} and {\em XMM-Newton}, its pulse period, $P$,  spin-up rate, $\dot P$, and luminosity, $L_{\rm X}$, are $\sim31.6\,\rm s$, $\sim-5.56\times10^{-7}\,\rm s\,s^{-1}$, and $\sim4.7\times10^{39}\,\rm erg\,s^{-1}$ in $0.3-30$ keV, respectively \citep{car18}. 
It has been spun up from $\sim32\,\rm s$ to $\sim20\,\rm s$ within two years after detecting the pulsation  \citep{bac18, vas18}. 
The field strength was derived to be {$ \sim(0.7-10)\times10^{12}\,\rm G$ \citep{vas18} and $\sim(3-20)\times10^{12}\,\rm G$ \citep{car18}} according to the accretion torque on the NS. 
And based on the accreting model for the magnetar with low magnetic field, it was deduced to be  $6.7\times10^{13}\,\rm G$ \citep{ton19}. 
Meanwhile, a strong field inferred by \citet{erk20} from the spin-up rate was $(45-240)\times10^{13}\,\rm G$.
In 2018, \citet{wal18} found a cyclotron resonant scattering feature (CRSF) in the spectrum of NGC 300 ULX1 at $E\sim13\,\rm keV$, that illustrated a $\sim10^{12}\,\rm G$ magnetic field if it was produced by electron. To account for the high spin-up rate, this CRSF was favored to be produced by proton and implied a magnetar field strength \citep{erk21}.  
However, \citet{kol19} doubted the existence of the CRSF. They introduced a model without the  additional absorption feature, which can describe the spectral and temporal characteristics of NGC 300 ULX1 successfully. Further detection for the CRSF of NGC 300 ULX1 is needed.  

Due to different results and the importance of the magnetic fields in understanding the characteristic of NGC 300 ULX1, in this work, we will study the field strength based on recent observations of the pulse period and spin-up rate.  In section~\ref{sec:calculation}, the dipole magnetic field of NGC 300 ULX1 is estimated through the torque model by \citet{gho79},  hereafter GL model. 
In section~\ref{sec: evolution}, we simulate the evolution of the magnetic field and spin period to inspect the rational field strength for NGC 300 ULX1. The evolved result is studied with the restrictions of the accretion rate and the companion mass.
Discussions and conclusions are given in section~\ref{sec: discussion} and~\ref{sec: conclusion}, respectively.

\section{The magnetic field strength calculation}
\label{sec:calculation}

\subsection{The magnetic field related to the accretion rate}

\begin{figure}
\centering
\includegraphics[width=8cm]{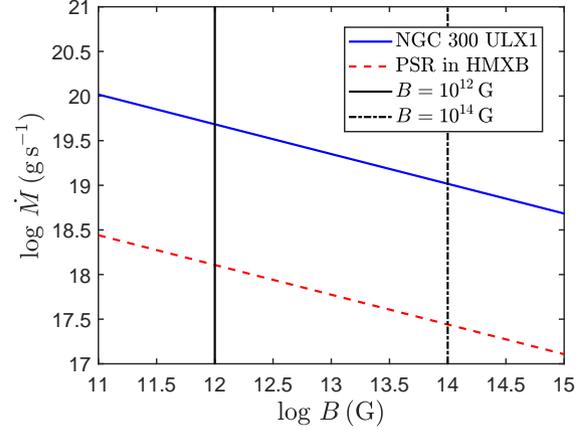}
\caption{Relation of $B$ and $\dot M$. Tracks of the normal PSR in HMXB and NGC 300 ULX1 are plotted in the red dotted and blue solid lines.}
\label{Md_B12}
\end{figure}

In an accreting binary system, the spin-up rate reflects the angular momentum being deposited on the NS \citep{sha83, ton15} 
\begin{equation}
I\dot{\Omega}=-2\pi I\frac{\dot{P}}{P^2}=\dot M \sqrt{GMR_{\rm A}},
\label{Iome}
\end{equation}
where $I$ is the moment of inertia of the NS, $\dot {\Omega}$ is the derivative of the angular velocity,  $\dot M$ is the accretion rate, $G$ is the gravitational constant, $M$ is the mass of NS, and $R_{\rm A}$ is the Alfv$\acute{\rm e}$n radius :
%
%
\begin{equation}
R_{\rm A}=3.5\times10^8m^{1/7} R_6^{-2/7}L_{37}^{-2/7}\mu_{30}^{4/7}\,\rm cm,
\label{ra}
\end{equation}
where $R_6$ is the NS radius $R$ in units of $10^6\,\rm cm$, $m$ is the NS mass in units of solar mass, $L_{37}$ is the accreting luminosity ($L_{\rm acc}=GM\dot{M}/R$) in units of $10^{37}\,\rm erg\,s^{-1}$, $\mu_{30}=1/2\,B_{12}R_6^3$ is the magnetic moment in units of $10^{30}\,\rm G\, cm^3$, $B_{12}$ is the dipole magnetic field in units of $10^{12}\,\rm G$.
With equations~(\ref{Iome}) and (\ref{ra}), the accretion rate can be written as    
\begin{equation}
\dot M^{6/7}_{18} \simeq-5.1\times10^{10}\,I_{45}m^{-3/7}
R_6^{-6/7}\,P^{-2}\dot{P}B^{-2/7}_{12}\,\rm g\,s^{-1},
\label{b_md}
\end{equation}
%
where $I_{45}$ is the moment of inertia $I$ in units of $10^{45}\,\rm g\,cm^2$. 
For a PSR with a certain spin period and spin-up rate, this equation shows an anti-correlation between the accretion rate and the magnetic field. 
We assume $P$ and $\dot P$ of a normal accreting PSR in high mass X-ray binary (HMXB) are $\sim 2\,\rm s$ and $\sim-1\times10^{-10}\,\rm s\,s^{-1}$, respectively \citep{liu06, liu07}. Its $B-\dot M$ relation according to equation~(\ref{b_md}) is shown in the red dotted line in Fig.~\ref{Md_B12}.
We see that the magnetic field $\sim10^{12}\,\rm G$ of such a PSR corresponds to the accretion rate  $\sim10^{18}\,\rm g\,s^{-1}$. If the magnetic field is as strong as $\sim10^{14}\,\rm G$, the accretion rate reduces to $\sim10^{17.5}\,\rm g\,s^{-1}$.  
And for NGC 300 ULX1, the $B-\dot M$ relation with the detected $P$ and $\dot P$ is plotted in the blue solid line in the same figure: if the magnetic field of NGC 300 ULX1 is $\sim10^{12}\,\rm G$, the accretion rate is about $\sim 10^{19.7}\,\rm g\,s^{-1}$.
When the magnetic field rises up to $\sim10^{14}\,\rm G$, the accretion rate is slightly higher than $\sim 10^{19}\,\rm g\,s^{-1}$.   
The two accretion rates correspond to the luminosity of $\sim9.4\times10^{39}\,\rm erg\,s^{-1}$ and $>\sim2.0\times10^{39}\,\rm erg\,s^{-1}$, where the later one is closed to the observed value of $4.7\times10^{39}\,\rm erg\,s^{-1}$. 
Thus we suggest that the magnetic field of NGC 300 ULX1 is about  $\sim10^{14}\,\rm G$, which will be investigated further in the following sections.

%

\subsection{The magnetic field deduced by the GL model}

During the accretion, the total torque $N$ on the NS is $N=I\dot{\Omega}$ for neglecting the typically small effect of the change in the effective moment of inertia.   
It can also be expressed in terms of the accretion torque $N_{\rm acc}$ \citep{gho79, gho95}:
\begin{equation}
N=n(\omega_{\rm s})N_{\rm acc}\simeq n(\omega_{\rm s})\dot M \sqrt{GMR_{0}},
\label{N}
\end{equation}
where $R_{\rm 0}$ denotes the transition zone, that is composed of a broad outer zone where the angular velocity is Keplerian and a narrow boundary layer where it departs significantly from the Keplerian value. It is preferred to be $\sim0.5R_{\rm A}$. The parameter $n(\omega_{\rm s})$ is the dimensionless torque related to the fastness parameter $\omega_{\rm s}$ \citep{gho79}
\begin{equation}
n(\omega_{\rm s})\simeq1.4\times(\frac{1-\omega_{\rm s}/\omega_{\rm c}}{1-\omega_{\rm s}}),
\label{n_ws}
\end{equation}
\begin{equation}
\omega_{\rm s}=1.19P^{-1}\dot M_{17}^{-3/7}\mu_{30}^{6/7}m^{-5/7},
\label{fastness}
\end{equation}
where $\dot M_{17}$ is $\dot M$ in units of $10^{17}\,\rm g\,s^{-1}$, 
$\omega_{\rm c}$ is the the critical fastness at which the torque changes direction and is taken to be 0.35 \citep{gho79, gho95}. In case of $\omega_{\rm s}<\omega_{\rm c}$, the PSR is spun up. And for $\omega_{\rm s}>\omega_{\rm c}$, the PSR is spun down \citep{gho95}.

According to equation~(\ref{N}), the spin-up rate is \citep{gho79} 
\begin{equation}
\label{pdot}
-\dot P\simeq 5.0\times10^{-5}\mu_{30}^{2/7} m^{-3/7}R_6^{6/7}I_{45}^{-1}P^2L_{37}^{6/7}n(\omega_s)~{\rm s~yr^{-1}}.
\end{equation}
In the follow calculation, we take the mass and radius of NGC 300 ULX1 to be the typical values of a standard NS, that are $m=1.4$ and $R_6=1$.

The pulse profile of NGC 300 ULX1 is found to be almost sinusoidal, and dose not appear to be strong beamed \citep{car18}. Moreover, the accretion rate of the source was detected to be  almost constant from 2016 to 2018 
\citep{vas19}. 
Thus we consider $L_{\rm acc}$ equals to $L_{\rm X}$ during the field strength calculation. 
Applying $P=31.6\,\rm s$ and $\dot P=5.56\times10^{-7}\,\rm s\,s^{-1}$ into equation~(\ref{pdot}), the magnetic field of NGC 300 ULX1 is confound when $\omega_{\rm s}$ is considered in two cases:\\
$\bullet$ $\omega_{\rm s}$ is assumed to be zero for a slow rotator, and the dimensionless torque is $n(\omega_{\rm s})\simeq1.4$, as defined by  equation~(\ref{n_ws})  \citep{gho95,par16}. The estimated magnetic field is {$B_1= 1.19\times10^{13}\,\rm G$, which is slightly stronger than} the works by \citet{car18} and \citet{vas18}.\\
$\bullet$ $\omega_{\rm s}$ is determined by equation~(\ref{fastness}) which is associated with the calculated magnetic field, spin period and luminosity. The inferred magnetic field is $B_2=3.89\times10^{14}\,\rm G$  and $\omega_{\rm s}\sim0.25$, that indicates NGC 300 ULX1 is a highly magnetized NS.

We list the inferred magnetic field $B$ and the corresponding Alfv$\acute{\rm e}$n radius $R_{\rm A}$ in Table~$\ref{single_B}$.
The co-rotation radius $R_{\rm co}$ is $1.68\times10^9\,\rm cm$ for $P=31.6\,\rm s$ according to $R_{\rm co}=(GM/4\pi^2)^{1/3}\,P^{2/3}$.
Both of the deduced magnetic fields by GL method imply the relations of $R_{\rm A}< R_{\rm co}$ and $\omega_{\rm s}<\omega_{\rm c}$. These results illustrate that the source is spinning up during the accretion, that is consistent with the observed case of NGC 300 ULX1.


More observations of NGC 300 ULX1 have been performed after it was confirmed as a PULX \citep{vas18}.  
Similar to the above calculation, we calculate magnetic fields by 15 pairs of $P$ and $\dot{P}$ that are taken from Table B.1 in \citet{vas18}, as can be seen in Table~\ref{More_B}.
\citet{vas18} suggested that the datasets, marked with the symbol ``*''  from the 2014 $\it Chandra$ and $\it Swift$/XRT observations, were with gaps and low statistics and therefore resulted in large uncertainties for determining the spin periods and spin-up rates.
The magnetic fields inferred from those observed parameters are with the same symbol ``*''. The mean value of the field solutions are $B_1\sim10^{13}\,\rm G$ with $n(\omega_{\rm s})\simeq1.4$ for the PSR as a slow rotator, and $B_2\sim3.0\times10^{14}\,\rm G$ with $n(\omega_{\rm s})$ according to equation~(\ref{n_ws}). 
In Table~\ref{More_B}, both $B_1$ and $B_2$ show a weak decay tendency in $3.4\,\rm yrs$ time span of the observation. In order to estimate a rational solution for NGC 300 ULX1, we perform a long term field decay analysis and compare the resulting field estimate with the magnetic field values in Table~\ref{More_B} inferred from observations.

%

\begin{table}
\centering
\caption{The magnetic field of NGC 300 ULX1 estimated with common parameters.}
\begin{tabular*}{\hsize}{@{}@{\extracolsep{\fill}}ccc@{}}
\hline
$B\,({\rm G})$ &
$\omega_{\rm s}$ &
$R_{\rm A}\,(\rm cm)$\\
\hline
$B_1=1.19\times10^{13}$&
$\sim0$&
$1.75\times10^8$\\
$B_2=3.89\times10^{14}$&
$\sim0.25$ &
$1.29\times10^9$\\
\hline
\end{tabular*}
\begin{flushleft}
{\bf Notes.} Magnetic fields strength are evaluated according to equation~(\ref{pdot}) with the employed parameters, that are $L_{\rm X}=4.7\times10^{39}\,\rm erg\,s^{-1}$, $P=31.6\,\rm s$ and $\dot P=-5.56\times10^{-7}\,\rm s\,s^{-1}$ \citep{car18}.
During the calculation, the field results depend on the dimensionless torque $n(\omega_{\rm s})$ associated with $\omega_{\rm s}$. The Alfv$\acute{\rm e}$n radius $R_{\rm A}$ are derived by the deduced magnetic fields. 
\end{flushleft}
\label{single_B}
\end{table}


\begin{table*}
\caption{The magnetic fields of NGC 300 ULX1 deduced with more observed properties}
\begin{center}
\begin{tabular}{llllllll}
\hline\noalign{\smallskip}
     \multicolumn{1}{c}{No.} &
     \multicolumn{1}{c}{Observatory/} &
     \multicolumn{1}{c}{\it {T$_{\rm obs}$}$^b$} &
     \multicolumn{1}{c}{\it P$_{\rm Zero}$ $^{c}$} &
     \multicolumn{1}{c}{\it $Log(|\dot{P}|)$ $^{d}$} &
     \multicolumn{1}{c}{$B_1$} &
      \multicolumn{1}{c}{$B_2$} &
      \multicolumn{1}{c}{$\omega_{\rm s}$}\\
     \multicolumn{1}{c}{} &
     \multicolumn{1}{c}{ObsID $^{a}$} &
     \multicolumn{1}{c}{MJD} &
     \multicolumn{1}{c}{s} &
     \multicolumn{1}{c}{$\rm s\,s^{-1}$} &
     \multicolumn{1}{c}{10$^{12}$ G}  &
      \multicolumn{1}{c}{10$^{14}$ G} &
      \multicolumn{1}{c}{}\\
     \noalign{\smallskip}
     \hline\noalign{\smallskip}
1&C-16029 & 56978.6 &126.28$^{(*)}$     & --4.94$^{(*)}$    & 
$29.16^{(*)}$ &  $0.31^{(*)}$ & $0.01^{(*)}$ \\
         \noalign{\smallskip}
2&S-00049834005 & 57502 &  44.18$^{(*)}$    & --6.0$^{(*)}$    & %
$8.86^{(*)}$ & $6.32^{(*)}$ & $ 0.28^{(*)}$ \\
         \noalign{\smallskip}
3&N-30202035002 & 57738 & 31.718      & --6.257    & 
11.36 & 3.93 & 0.26 \\
        \noalign{\smallskip}
4&X-0791010101 & 57739 & 31.683  & --6.25  & 
12.12 &  3.89 &  0.25 \\
        \noalign{\smallskip}
5&X-0791010301 & 57741 & 31.588  & --6.25  & 
12.37  & 3.86 &  0.25 \\
        \noalign{\smallskip}
6&S-00049834008 & 57860 &  26.87$^{(*)}$ & --6.34$^{(*)}$    & %
$18.59^{(*)}$ & $2.80^{(*)}$ & $ 0.23^{(*)}$ \\
         \noalign{\smallskip}
7&S-00049834010 & 57866 & 26.65$^{(*)}$ &  --6.4$^{(*)}$     & %
$12.14^{(*)}$ & $3.06^{(*)}$ & $ 0.24^{(*)}$ \\
        \noalign{\smallskip}
8&S-00049834013 & 57941 & 24.24$^{(*)}$ & --6.8$^{(*)}$      & %
$ 0.94^{(*)}$  & $3.54^{(*)}$ & $ 0.31^{(*)}$ \\
        \noalign{\smallskip}
9&S-00049834014 & 57946 & 24.22$^{(*)}$ &  --6.5$^{(*)}$      & %
$ 10.59^{(*)}$ & $ 2.75^{(*)}$ &$ 0.25^{(*)}$ \\
        \noalign{\smallskip}
10&S-00049834015 & 58143 & 20.06$^{(*)}$ &  --6.5$^{(*)}$      & %
$ 1.58^{(*)}$  & $ 2.71^{(*)}$ & $ 0.29^{(*)}$\\
        \noalign{\smallskip}
11&N-90401005002 & 58149 & 19.976 & --6.74    &
5.90 & 2.34 & 0.26 \\
        \noalign{\smallskip}
12&C-20965 & 58157 & 19.857 & $<-6.7$ & 
$<8.43$  & $>2.18$ & $>0.25$ \\
         \noalign{\smallskip}
13&C-20966 & 58160 & 19.808 &  $<-6.7$ & 
$<8.64$  & $>2.16$ & $>0.24$ \\
         \noalign{\smallskip}
14&C-20965/20966 & 58157 & 19.857&  --6.82  & 
$3.23$ & $ 2.51$ &  0.28 \\
         \noalign{\smallskip}
15&S-00049834019 & 58221 & 19.046$^{(*)}$ &  --6.6$^{(*)}$   & %
 $25.44^{(*)}$  & $1.34^{(*)}$ & $0.17^{(*)}$ \\
        \noalign{\smallskip}
\hline\noalign{\smallskip}
\end{tabular}
\newline
{\bf Notes.}
$^{(a)}$ Observation ID for {\it XMM-Newton} (X), {\it Chandra} (C) and {\it Swift} (S).
$^{(b)}$ $T_{\rm obs}$: start day of observation.
$^{(c)}$ Pulse period.
$^{(d)}$ Spin-up rate.
Magnetic fields inferred from more observed parameters of $T_{\rm ac}$, $P_{\rm Zero}$ and $Log(|\dot{P}|)$, that are given by \citet{vas18}. The calculated $B_1$ is in case of $n(\omega_{\rm s})\sim1.4$ for $\omega_{\rm s}\sim0$ and $B_2$ is with $n(\omega_{\rm s})$ associated with $\omega_{\rm s}$ defined by equation~(\ref{fastness}). Datas marked with symbol ``*'' can not be determined a unique set of values due to aliasing and multiple possible solutions caused by gaps and low statics of observations. And we mark
the deduced magnetic fields with the same symbol.
\end{center}
\label{More_B}
\end{table*}

\section{The evolved simulation}
\label{sec: evolution}

\subsection{Evolution of the magnetic field}

Evolution of NS magnetic field during the accretion has been widely studied and discussed \citep{bis74, gep94, heu95, mel00, heu09, ho11, igo15, igo21}. For estimating a reasonable magnetic field of NGC 300 ULX1 from the inferred solutions, we will simulate the magnetic field decay along with the accretion time $T_{\rm ac}$ by employing  the model of the accretion induced the magnetic field decay of a NS by \citet{zha06}, hereafter ZK model. It had been used to test the magnetic field evolution of binary pulsars (BPSRs) \citep{pan15}.

The ZK model assumes that the PSR is evolved with a constant accretion rate in the accretion. Since the accreted material flows from the polar to the equator, the magnetic field lines in the polar caps are pushed aside and the polar-field strength are diluted. As the PSR magnetosphere is compressed onto the star surface with the accumulation of the accretion material, the accretion might become all over the star and the magnetic field of the PSR decays to the bottom field. The analytic solution of the evolved magnetic field is \citet{zha06}:
\begin{equation}
\label{zk06}
B=\frac{B_{\rm f}}{\{1-[C/{\rm exp}(y)-1]^2\}^{7/4}},
\end{equation}
where $B_{\rm f}\simeq 1.32\times10^8(\dot M/\dot M_{\rm Edd})^{1/2}m^{1/4}R_6^{-5/4}\phi^{-7/4}\,\rm G$, is the bottom magnetic field, $\dot M_{\rm Edd}\simeq1.0\times10^{18}\,\rm g\,s^{-1}$ is the typical Eddington accretion rate of a NS with a mass of $1.4M_\odot$ and radius of 10 km, $\phi$ is assumed to be 0.5, denoting the ratio of the magnetosphere radius to the Alfv$\acute{\rm e}$n radius.
$C=1+[1-(B_{\rm f}/B_0)^{4/7}]^{1/2}$, $B_0$ is the initial magnetic field when the PSR begins the evolution. The parameter $y=2\xi\Delta M/7M_{\rm cr}$, where $M_{\rm cr}\sim0.2\,M_{\odot}$ is the crust mass, $\Delta M=\dot M T_{\rm ac}$ is the accreted material. The parameter $\xi$ $(0\leq\xi\leq1)$ is the efficiency factor to express the frozen flow of the magnetic line due to the plasma instability. In this work, we assume $\xi=1$ for the completely frozen field lines that totally drift with the accreted material. The model illustrates that the field evolution of the NS is mainly affected by $\dot M$ and $T_{\rm ac}$.

\begin{figure*}
\centering
$\begin{array}{c@{\hspace{0.1in}}c}
\multicolumn{1}{l}{\mbox{}} & \multicolumn{1}{l}{\mbox{}} \\
\includegraphics[width=\columnwidth]{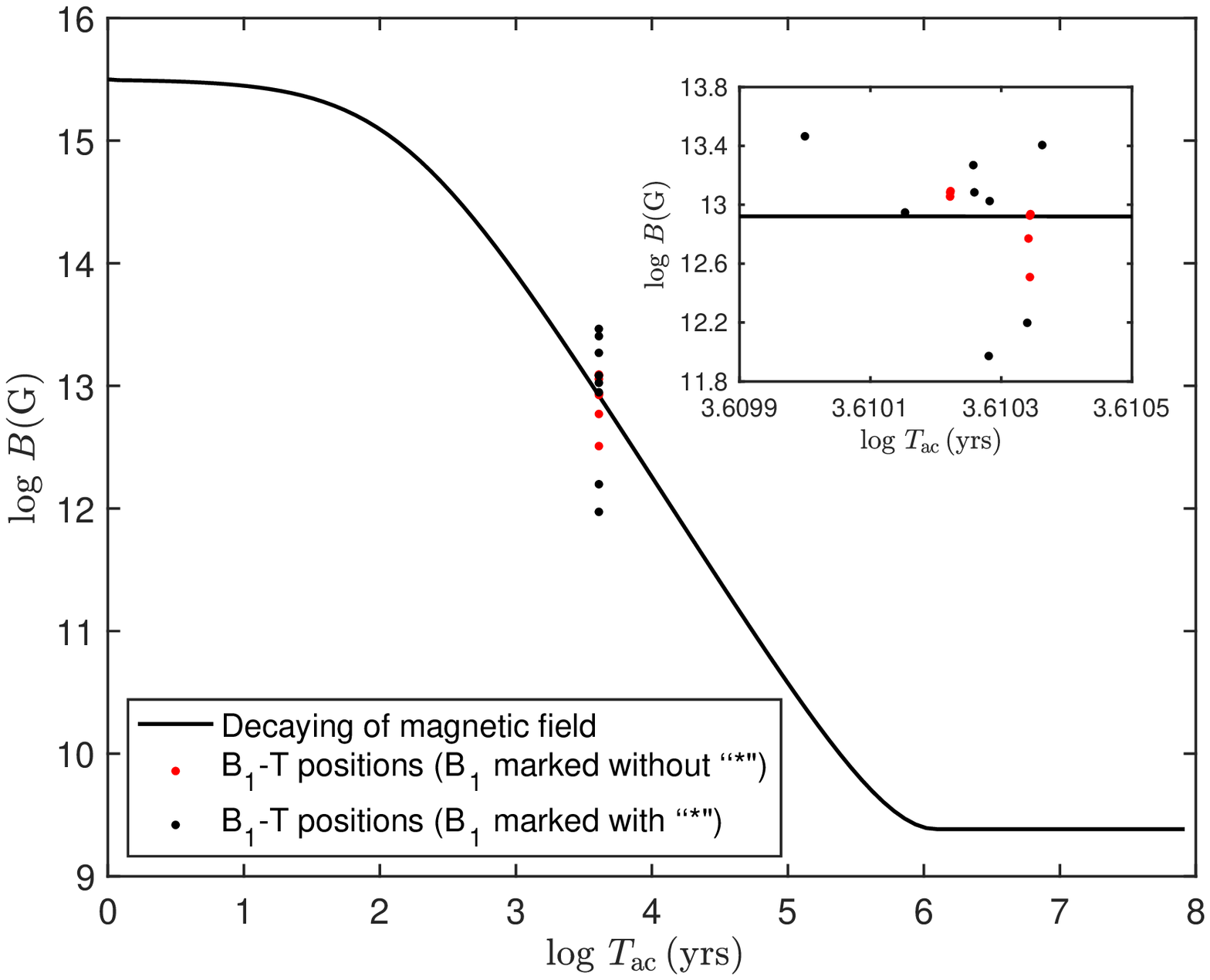} &\includegraphics[width=\columnwidth]{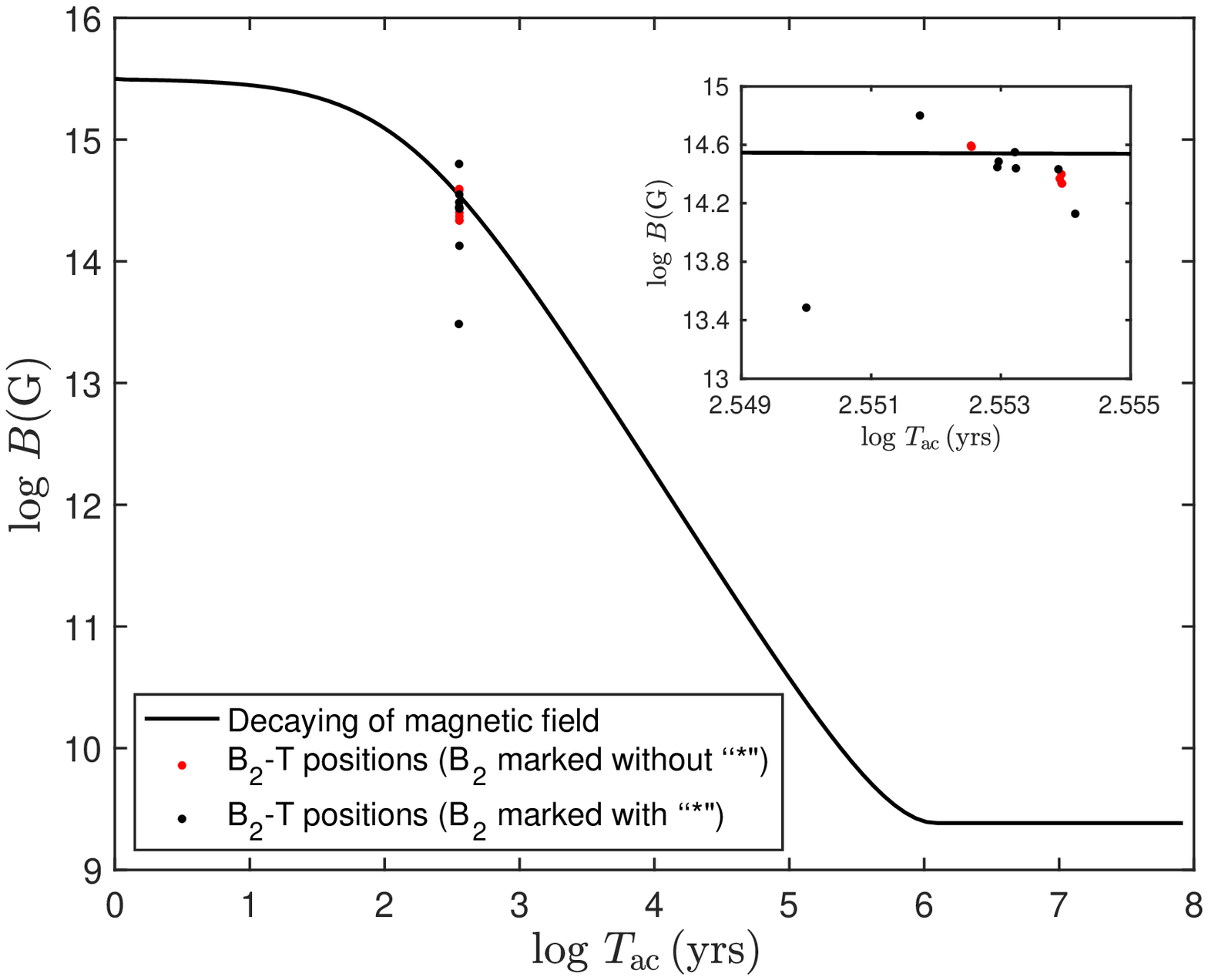}\\
\end{array}$
\caption{Decaying of the magnetic field of NGC 300 ULX1. We assume that the evolution begins with $T_{\rm ac}=0$ and $B_0\sim10^{15.5}\,\rm G$ in two panels. The tracks plotted with the black-solid lines show the the magnetic field decays to the bottom value of $\sim10^{9}\,\rm G$ within about $10^7\,\rm yrs$. The $B-T_{\rm ac}$ positions are plotted with the derived values of $B_1$ and $B_2$ in Table~\ref{More_B}, where the black and red dots are the fields marked with and without symbol  ``*'', respectively. The enlarged images of the PSR positions to evolved tracks 
are embedded in two panels.}
\label{BT_evolve}
\end{figure*}

During the evolution of $B$ with $T_{\rm ac}$, the initial magnetic field $B_0$ of NGC 300 ULX1 should be equaled to or stronger than the deducted magnetic field that are listed in Table \ref{More_B}. And $\dot M$ is assumed to be a constant corresponding to $L_{\rm X}=4.7\times10^{39}\,\rm erg\,s^{-1}$. It should be noted that the observation time does not mean the accretion time. Hence, we plot the $B-T_{\rm ac}$ relation to find part of the evolved track, which has a similar distribution to the observed $B$ in a time span of $3.4\,\rm yrs$. After testing different values of $B_0$, two $B-T_{\rm ac}$ tracks with $B_0\sim3.0\times10^{15}\,\rm G$ for $B_1$ and  $B_2$ are explored, that couple with the PSR positions well. The accretion time for the source field decaying from $B_0$ to the deduced value are $\sim4000\,\rm yrs$ for $B_1$ and $\sim400\,\rm yrs$ for $B_2$, as shown in the $B-T_{\rm ac}$ diagram in Fig.~\ref{BT_evolve}. The left panel displays the evolved track and the PSR positions with $B_1$, and the right panel reveals the ones with $B_2$. The sub-figure in each panel is the enlarged image for showing the detailed matching of the evolved track and positions. The red and black solid dots are the $B-T_{\rm ac}$ positions whose magnetic fields are with and without symbol "*" in Table~\ref{More_B}, respectively. We see that the $B-T_{\rm ac}$ positions in the right panel are on or closed to the evolved track more tightly than that in the left panel. Moreover, the reduced chi-square of the simulated and calculated magnetic field are $\sim45$ for $B_1$ and $\sim{\bf 2}$ for $B_2$, which also supports the higher magnetic field of NGC 300 ULX1. So we propose that the magnetic field of NGC 300 ULX1 would be $\sim3\times10^{14}\,\rm G$ rather than $\sim10^{13}\,\rm G$.  

\subsection{The evolved results}

\begin{figure}
\centering
\includegraphics[width=\columnwidth]{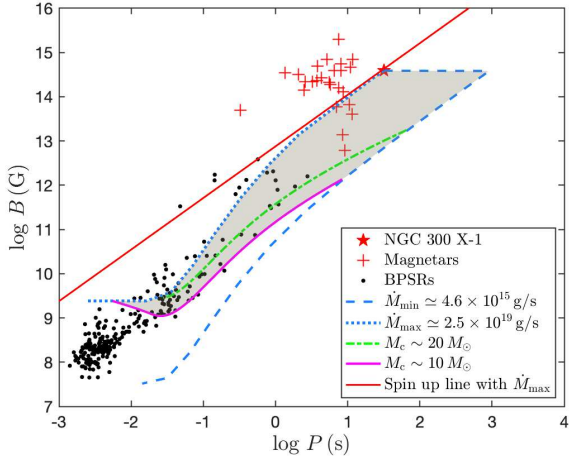}
\caption{B-P evolved tracks of NGC 300 ULX1. The tracks with the constant accretion rates of $\dot M_{\rm max}\simeq2.5\times10^{19}\,\rm g\,s^{-1}$ and $\dot M_{\rm min}\simeq4.6\times10^{15}\,\rm g\,s^{-1}$ are plotted with the blue dotted and blue dashed lines. The area in grey is the evolutionary range of the source, which is limited by the evolved tracks determined by the variate accretion rate and the companion mass (green dash-doted line for $M_{\rm c}=20\,M_{\odot}$ and magenta solid line for $M_{\rm c}=10\,M_{\odot}$). Red star is the B-P position of NGC 300 ULX1 with $B= 3.89\times10^{14}\,\rm G$ and $P=31.6\,\rm s$. Crossings and solid dots mean the B-P positions of the radio-loud magnetars and the binary pulsars, respectively.}
\label{BP_evolve}
\end{figure}

With the preferred magnetic field strength $B\sim3.0\times10^{14}\,\rm G$ of NGC 300 ULX1, the B-P evolution is studied according to equations~(\ref{pdot}) and (\ref{zk06}).
We first consider a simple case that the source has been experiencing a steady accretion with the defined accretion time and rate. 
The accretion time is assumed to be the Hubble time $1.38\times10^{10}\,\rm yr$ \citep{pla16}.  
And the accretion rate is with two limitations: the maximum value corresponds to its  ultra luminosity $L_{\rm X}=4.7\times10^{39}\,\rm erg\,s^{-1}$.
And the minimum value is $\dot M_{\rm min}\simeq4.6\times10^{15}\,\rm g\,s^{-1}$, which is also the minimum accretion rate required for a MSP forming in the binary system \citep{pan18}. The corresponding bottom magnetic fields are $B_{\rm f,max}\simeq1.73\times10^9\,\rm G$ and $B_{\rm f,min}\simeq2.34\times10^7\,\rm G$, respectively. Under these conditions, the B-P evolution began with $B\simeq3.0\times10^{14}\,\rm G$ and $P=31.6\,\rm s$ are shown in Fig.~\ref{BP_evolve}, where the blue dotted and dashed lines are with $\dot M_{\rm max}$ and $\dot M_{\rm min}$. We find that NGC 300 ULX1 will evolve to be a MSP if the source kept accreting throughout the Hubble time.

Since it is hard for a PSR to keep a constant accretion rate during the entire evolution, we now consider that the accretion rate of NGC 300 ULX1 is  changeable and fluctuates between $\dot M_{\rm max}$ and $\dot M_{\rm min}$. And the source will evolve to different equilibrium status along with the accretion rate variation \citep{ton19, bha91, ho14}. The accretion time $T_{\rm ac}$ is restricted by the companion mass, which can be inferred from 
\begin{equation}
T_{\rm ac}=1.3\times10^{10}\, f \,m_{\rm c}^{-2.5}\,\rm yr,
\label{mc_lf}
\end{equation}
where $m_{\rm c}$ is the companion mass in units of solar mass, $f$ is the accreted efficient factor that is usually taken to be 0.1 \citep{sha83}.
It has been detected NGC 300 ULX1 is with a $<20M_{\odot}$ companion \citep{car18, bin20}. And we consider a lower mass limit of the companion to be $\sim10M_{\odot}$ as it is in HMXB. The corresponding accretion time is $\sim10^6-10^7\,\rm yrs$ according to equation~(\ref{mc_lf}). with the variation of the accretion rate and the accretion time, two evolved B-P tracks are shown in Fig.~\ref{BP_evolve}, where the magenta solid line and green dot-dashed line are plotted with $T_{\rm ac}\sim10^6-10^7\,\rm yrs$. Now we see that NGC 300 ULX1  evolves to a MSP whose spin period is $\sim10\,\rm ms$.

Restricted by three B-P tracks that are the blue dotted line, blue dashed line and magenta solid line (for $M_{\rm c}=10\,M_{\odot}$) or green dot-dashed line (for $M_{\rm c}=20\,M_{\odot}$), an evolved range of NGC 300 ULX1 is got, that is covered in grey in Fig.~\ref{BP_evolve}. It shows all evolved possibilities of NGC 300 ULX1, e.g., a MSP with $B\sim10^9\,\rm G$, or an evolved PSR with $P$ about tens of seconds and $B\sim10^{10}\,\rm G$.

\section{Discussions}
\label{sec: discussion}

\subsection{Comparison with other works}

By the $B-\dot M$ relation and the GL model, we deduced the magnetic field of NGC 300 ULX1 with the X-ray luminosity, 15 pairs of the observed spin periods and spin-up rates. During the calculation with the GL method, we find that the magnetic field is strongly depends on the fastness parameter $\omega_{\rm s}$ in the GL model. For $\omega_{\rm s}$ defined by equation~(\ref{fastness}), $B\sim 3.0\times10^{14}\,\rm G$. It is consistent with the work by Erkut et al. (2020), who estimated the maximum range of the magnetic dipole field to be $(45-240)\times10^{13}\,\rm G$ from the spin-up rate. And in case of a slow rotator ($\omega_{\rm s}\sim0$),  $B\sim10^{\bf 13}\,\rm G$, that is closed to the results of $\sim (0.7-10)\times10^{12}\,\rm G$ by \citet{car18} and $\sim (3-20)\times10^{12}\,\rm G$ by \citet{vas18}.

The field strength $\sim10^{13}\,\rm G$ of NGC 300 ULX1 is excluded according to the  evolution of the magnetic field decaying with the accretion time and the chi-square test for the simulated and the calculated magnetic fields. Based on these study, we conclude that the magnetic field of NGC 300 ULX1 is $\sim3.0\times10^{14}\,\rm G$, and suggest the fastness parameter $\omega_{\rm s}$ to be a variate instead of a constant during the magnetic field deduction.

\subsection{Magnetic field constrained by CRSF}

CRSF is a direct way for constraining the magnetic field of a PSR. In the confirmed PULXs, M51 ULX8 is the only source that was detected with a certain absorption feature at $\sim4.5\,\rm keV$. It was preferred to be with a proton CRSF which implied a magnetic field of $\sim10^{15}\,\rm G$ \citep{bri18}. NGC 300 ULX1 is the second PULX that was probed a potential CRSF at $\sim13\,\rm keV$ based on the phase-resolved broadband spectroscopy using the data of $\it XMM-Newton$ and $\it NuSTAR$, which implied the magnetic field to be $\sim10^{12}\,\rm G$ with the assumption of scattering by electrons \citep{wal18}. The recent work favored a proton CRSF that could lead to a magnetar field strength of NGC 300 ULX1. It was consistent with the high spin up rate of the source to estimate the plausible ranges for the beaming fraction \citep{erk21}.

Meanwhile, such a CRSF needed a further confirmation, since \citet{kol19} found a model without the absorption feature could successfully describe the spectra of NGC 300 ULX1. Therefore, the magnetic field of NGC 300 ULX1 has to be derived through the theory models, such as the accretion torque or the model of the magnetic field decay by accretion. And We look forward to an undoubtable detection of CRSF to determine the magnetic field of NGC 300 ULX1 directly.

\subsection{Relations to magnetars}

Limited by the accretion rate and accretion time, NGC 300 ULX1 will evolve to an recycled PSR, whose magnetic field is $\lesssim10^{12}\,\rm G$ and the spin period is $\lesssim10\,\rm s$, as shown with the grey area in the B-P diagram in Fig. \ref{BP_evolve}. We plot the B-P positions of magnetars and binary PSRs in the same figure, whose data are from McGill Online Magnetar Catalog and ATNF pulsar catalogue \citep{man05, ola14}. It is found that: (1) the present position of NGC 300 ULX1 is closed to the magnetars, (2) its possible B-P evolved area covers not only the magnetars with low magnetic field ($\sim10^{12}\,\rm G$), but also the gap between magnetars and BPSRs.

It is known that the magnetars are always found as the isolated PSRs whose magnetic field are $\sim10^{14}-10^{15}\,\rm G$ \citep{ola14,pop16, kas17}. And there were also a few magnetars whose magnetic field is slightly weaker $\sim10^{12}\,\rm G$, that were suspected to undergo the accretion induced their fields decay \citep{rea10,zho14, mer15, esp21}. Meanwhile, there is no binary magnetar found until now. Although an accreted magnetar was doubtful existed in the $\gamma$-ray binary system LS 5039 with a period of $\sim9\,\rm s$ \citep{yon20}. It was denied later since the statistical significant bursts or quasiperiodic variability was not found with the same data \citep{vol21}. If NGC 300 ULX1 is the accreting magnetar, it might provide us a chance for probing the magnetar evolution and explaining the formation of the magentars with the low magnetic fields.

\subsection{Ultra X-ray luminosity}

Our work suggests that NGC 300 ULX1 is an accreting magnetar whose magnetic field is about  $3.0\times10^{14}\,\rm G$, which would be with even more strong multi-pole magnetic field. The strong magnetic field can reduce the electron scattering cross-section of the PSR, and in turn supports the super-critical accretion and ultra X-ray luminosity to occur.

The ultra X-ray luminosity of PULXs can be accounted for by the beaming effect \citep{kin20, son20}. However, some confirmed PULXs showed nearly sinusoidal pulse profile which means the beaming effect is negligible. \citet{mus21} also suggested that the large pulsed fraction of PULXs excluded the strong beaming \citep{eks15, fur17,car18}. Since NGC 300 ULX1 was observed with a similar property \citep{car18}, we propose that the accretion luminosity is indeed  supper-Eddington.

\section{Conclusions}
\label{sec: conclusion}

In this paper, we study the magnetic field strength of NGC 300 ULX1 to be $\sim3.0\times10^{14}\,\rm G$ with the assumption that the source is undergoing a constant accretion rate. Since the pulse profile of NGC 300 ULX1 is almost sinusoidal, the beaming effect is ignored. The B-P evolved simulation, which is confined by the accretion time and accretion rate, shows the source can evolve to a recycled PSR, for instance, a MSP with $B\sim10^{9}\,\rm G$ or an evolved BPSR with $P<10\,\rm s$ and $B\sim10^{10}\,\rm G$.

Taking into account the resulting $ 3.0\times10^{14}\,\rm G$ field following the accretion induced decay of an initial $\sim 3.0\times10^{15}\,\rm G$ field, we propose NGC 300 ULX1 is probably an accreting magnetar. We also suggest there might be other magnetars among other PULXs such as M82 X-2, NGC 7793 P13 and NGC 5907 ULX \citep{pan16, isr17, fur17, ton19}. These PULXs can be probed to study the magnetar evolution and the MSP formation in binary systems.

\section*{Acknowledgements}

We would like to thank Liming Song, Na Wang and Qingzhong Liu for helpful discussions. YP, ZL and CZ are supported by National Natural Science Foundation of China (12130342, U1938107, U1838111, U1838117).

\section*{Data Availability}

The data underlying this article will be shared on reasonable request to the corresponding author.



\bibliographystyle{mnras}
\bibliography{NGC300_refs} 

\bsp	
\label{lastpage}
\end{document}